\documentclass{article}
\pdfoutput=1

\usepackage{booktabs} 

\usepackage{amsmath,amssymb,enumerate,comment, color,lipsum,url}
\usepackage{graphicx,caption,subcaption, latexsym, tkz-graph, color, float, tikz,framed}
\usepackage{pseudocode, array, numprint, verbatim}
\usepackage[normalem]{ulem}
\usepackage[linesnumbered,ruled]{algorithm2e}

\usepackage{authblk}

\def \etal { \emph{et~al.} }

\newcommand{\name}{CADDeLaG\xspace}
\SetKwProg{Fn}{Function}{}{}

\title{More than one Author with different Affiliations}
\author[1]{Aniruddha Basak\thanks{abasak@cmu.edu}}
\author[2]{Kamalika Das\thanks{kamalika.das@nasa.gov}}
\author[3]{Ole J. Mengshoel\thanks{ole.mengshoel@sv.cmu.edu}}
\affil[1,3]{Carnegie Mellon University}
\affil[2]{NASA Ames Research Center}

\begin{document}
\title{CADDeLaG: Framework for distributed anomaly detection in large dense graph sequences}


%
%
%
%
%

\begin{abstract}
Random walk based distance measures for graphs such as commute-time distance are useful in a variety of graph algorithms, such as clustering, anomaly detection, and creating low dimensional embeddings. Since such measures hinge on the spectral decomposition of the graph, the computation becomes a bottleneck for large graphs and do not scale easily to graphs that cannot be loaded in memory. Most existing graph mining libraries for large graphs either resort to sampling or exploit the sparsity structure of such graphs for spectral analysis. However, such methods do not work for dense graphs constructed for studying pairwise relationships among entities in a data set. Examples of such studies include analyzing pairwise locations in gridded climate data for discovering long distance climate phenomena. These graphs representations are fully connected by construction and cannot be sparsified without loss of meaningful information.  In this paper we describe \name, a framework for scalable computation of commute-time distance based anomaly detection in large dense graphs without the need to load the entire graph in memory. The framework relies on Apache Spark's memory-centric cluster-computing infrastructure and consists of two building blocks: a decomposable algorithm for commute time distance computation and a distributed linear system solver. We illustrate the scalability of \name and its dependency on various factors using both synthetic and real world data sets. We demonstrate the usefulness of \name in identifying anomalies in a climate graph sequence, that have been historically missed due to ad-hoc graph sparsification and on an election donation data set.
\end{abstract}

%
%
\maketitle

\section{Introduction}

Graph based analysis of data sets facilitates identification of interesting relationships among data entities that may be otherwise missed when studied in a non-relational setting. Graphs representing human interactions such as social and organizational networks, telecommunication networks, collaboration networks, etc. are examples of data sets where entity relationships are an essential descriptor of the data. These graphs are often large and sparse. However, an entity-entity relationship needs to be induced in the data set through the use of a similarity measure when it becomes essential to study one entity in the context of others. For example, in order to study changes in the global precipitation pattern using time series observations of precipitation at different locations on the Earth, it may be more beneficial to study changes in pairwise relationships of those locations over time, rather than the independent observations \cite{sricharan2014localizing}. This requires constructing fully connected temporal graphs with nodes and edges representing locations and their similarities respectively. Studying the changes in the graph structure for such graph sequences can reveal important changes in precipitation patterns over time. Unfortunately, these graphs are dense by virtue of construction and the existing spectral methods for large sparse graphs do not apply here. For a spatial resolution of $0.5^\circ$, there are approximately 260,000 geolocations leading to a graph with 260,000 nodes and ($260,000 \times 260,000$) 67,600,000,000 edges. Such graphs are impossible to analyze using centralized computation due to memory requirements. Sparsification based on thresholds can often lead to loss of relevant information. Spectral sparsification of graphs have shown promise in graph structure analysis, and there are efficient computations for it \cite{spielman2011graph}.

To address the issue of graph structure analysis using commute time embedding, we propose \name: \underline{\textbf{C}}ommute-time based \underline{\textbf{A}}nomaly \underline{\textbf{D}}etection in \underline{\textbf{De}}nse \underline{\textbf{LA}}rger-than-memory \underline{\textbf{G}}raphs, a framework for scalable computation of commute-time distance based anomaly detection in large dense graphs without the need to load the entire graph in memory. The framework relies on Apache Spark's memory-centric cluster-computing infrastructure and consists of two building blocks: a decomposable algorithm for commute time distance computation and a distributed linear system solver for symmetric diagonally dominant (SDD) matrices. The foundation of the SDD solver is built upon inverse chain-based approximation \cite{peng2014efficient}. Parallelization of the chain computation, however,  needs loading of the matrix in memory. An input/output (I/O) and memory optimized Spark-based implementation of the matrix-chain computation algorithm for larger-than-memory matrices in a distributed memory cluster computing environment is in the heart of our proposed anomaly detection framework for large dense graphs. We illustrate the scalability of \name and its dependency on various factors using both synthetic and real world data sets. We demonstrate the usefulness of \name in identifying anomalies in a climate graph sequence, that have been historically missed due to ad-hoc graph sparsification. Experimental results on the 2016 presidential election donation data set indicate that our graph structural anomaly detection algorithm has the ability to capture the shifts in donor sentiment even when standard non-relational data analyses fail to do so.

The rest of this paper is structured as follows.  In section \ref{background} we present a brief description of CAD \cite{sricharan2014localizing}, an algorithm that successfully uses commute time distance for anomaly detection in graphs and approximates the computation using a linear SDD solver. Then in Section \ref{caddelag} we describe the details of our  proposed framework. Section \ref{expmts} shows the scalability analysis of our framework for large dense graphs. We report the results of anomaly detection using \name in Section \ref{anomalies}. We finally conclude the paper with some future research directions in Section \ref{conclusion}.


\section{Background}\label{background}
Let $G_t$, $t=1,\ldots,T$ be a temporal sequence of graphs. Each $G_t$ is a weighted undirected graph with a fixed vertex (node) set $V=\{v_1,..,v_n\}$ with an edge set $E=\{e_{(1,1)},..,e_{(n,n)}\}$ of size $m = n^2$, with $e_{(i,j)}$ denoting the edge between nodes $v_i$ and $v_j$. For each graph $G_t$, the symmetric adjacency matrix is denoted by $A_t \in \mathbb{R}^{n \times n}$, \emph{i.e.} the edge weight of edge $e_{(i,j)}$ in graph $G_t$ is given by $A_t(i,j)$. The graph does not have self-edges, therefore $A_t(i,i) = 0 \forall i$. Given any adjacency matrix $A$, if we can arbitrarily orient the edges, we can write the graph Laplacian matrix $L$ as \cite{spielman2011graph}
\begin{align}
L = B^TWB
\end{align}
where $B \in \mathbb{R}^{m \times n}$ is the signed edge-vertex incidence matrix,
\begin{align}
B(e,v) = \begin{cases}
1 \quad \text{if } v \text{ is } e \text{'s head} \\
-1 \quad \text{if } v \text{ is } e \text{'s tail} \\
0 \quad \text{otherwise } 
\end{cases}
\label{eq:def-B}
\end{align}
\noindent and $W \in \mathbb{R}^{m \times m}$ is the diagonal matrix with $W(e,e) = A(e_i, e_j)$. The commute time is the expected number of steps that a random walk starting at $i$ will take to reach $j$ once and go back to $i$ for the first time~\cite{khoa2010robust}. The commute time can be computed from the Moore-Penrose pseudoinverse of $L$ as  
\begin{equation}
c(i,j) = V_G(l^+_{ii}+l^+_{jj}-2l^+_{ij}),
\label{eq1}
\end{equation}
\noindent where $V_G = \sum_{i=1}^{n}D(i,i)$ is the volume of the graph, and $l^+_{ij}$ is the  $(i,j)$ element of inverse Laplacian $L^+$. Commute time distance has been proven to successfully capture structure in graphs  for clustering~\cite{khoa2010robust}, and for  anomaly detection in dynamic graphs~\cite{sricharan2014localizing}.

Commute time distance between the nodes of a graph is the Euclidean distance in the space spanned by the eigenvectors of the graph Laplacian $L$ corresponding to $A$. Therefore, computation of commute time distance for a single graph instance having $n$ nodes using Eqn. \eqref{eq1} is $O(n^3)$, which is prohibitively expensive for large graphs with several thousands of nodes. To deal with this problem, Khoa and Chawla proposed the use of approximate commute time embedding \cite{khoa2012large} that does not require eigen decomposition of the graph matrix. Instead it relies on the idea of spectral sparsification through random projection proposed by Speilman and Srivastava \cite{spielman2011graph} that can be solved using a linear time solver for SDD systems~\cite{spielman2014nearly}.

Anomaly localization in graphs refer to the problem of identifying abnormal changes in node relationships (edges) that cause anomalous changes in the graph structure. The commute time based anomaly detection algorithm  CAD \cite{sricharan2014localizing} has been found to be useful for localizing anomalies in temporal graph sequences. The CAD algorithm's goal is to detect, for each transition between graph instances $G_t$ and $G_{t+1}$,  the set of anomalous edges $E_t \subseteq E$ whose change in weights are {{responsible}} for structural differences between $G_t$ and $G_{t+1}$. CAD achieves this by computing the commute time distances $c_t(i,j)$ for every pair of nodes $v_i,v_j \in V$ and every time instance $t=1,..,T$ with an embedding dimension $k$ using \cite{khoa2012large}. The final score for each edge $\Delta E_t(.)$ given by
\begin{equation}
\Delta E_t(e_{i,j}) = |A_{t+1}(i,j) - A_t(i,j)| \times |c_{t+1}(i,j) - c_t(i,j)| \nonumber
\end{equation}
is computed for each graph transition and the top anomalous nodes are the ones for which $\sum_{j}{\Delta E_t(i,j)}$ are the highest.

\subsection{Related Work}\label{relwork}

Many real-world data-mining applications benefit from relational, network, or graph analysis: spectral clustering; graph anomaly detection; and creating low-dimensional graph embeddings.   Spectral clustering is powerful, but due to computational cost researchers adopt numerical approximation, sampling or spectral sparsification approaches. Fowlkes et al.~used the Nystrom technique to approximate solution of spectral partitioning of image and video~\cite{fowlkes04spectral}. Wang et al.~use column sampling to solve the eigensystem in a smaller sample and extrapolate the solution to the whole data set~\cite{wang09approximate}. Spielman and Teng introduced graph sparsification using spectral similarity of graph Laplacians~\cite{spielman2011spectral}. Effective resistance is also used for graph sparsification~\cite{spielman2011graph}.

Our work belongs to the category of graph sparsification. The previously mentioned approaches require storing the graph in memory and therefore is limited upto moderate size graphs. Tutunov et al.~have developed a distributed solver for SDD systems \cite{tutunov2015fast}.  Their approach, which is based on the parallel SDD linear systems solver by Peng and Spielman \cite{peng2014efficient},  uses an inverse chain approximation.  The distributed SDD solver is very complex to implement and no source code is publicly available. Christina \etal proposed an asynchronous solution to solve SDD systems using cloud computing model, but it only applies to sparse graphs where the local neighborhood of any node is bounded~\cite{lee2014solving}. Khoa and Chawla have developed an incremental ({\em i.e.}, online) approach to commute time computation based on streaming~\cite{khoa2016incremental}. However, the assumptions of the incremental approach only applies to streaming graphs where the structure does not change drastically, and thus does not apply for general anomaly detection. 

We discuss spectral methods in the context of graph anomaly detection~\cite{akoglu15Graph} and our work focuses on dynamic graphs which has numerous applications, for example in intrusion detection, telecommunication networks, and social networks~\cite{ranshous15anomaly}. We are primarily interested in localization of anomalies in dynamic graphs. Other approaches to anomaly localization  are OddBall by Akoglu et al.~\cite{akoglu10oddball}, the work by Wang et al.~\cite{wang15localizing}, and DeltaCon by Koutra et al. \cite{ koutra13deltacon}.

\section{CADD\lowercase{E}L\lowercase{A}G Framework}\label{caddelag}

In this paper we propose to extend the commute time based anomaly detection algorithm CAD \cite{sricharan2014localizing} for large dense graphs that do not fit in memory. The major compute intensive part of CAD is to compute the commute time distances for every pair of nodes in the two graphs $G_t$ and $G_{t+1}$. The computation is approximate where the amount of approximation error $\epsilon_{RP}$ is controlled by the random projection dimension $k_{RP}$. Algorithm \ref{commutetime_algo} describes Spielman and Srivastava's near-linear time algorithm for computing approximate effective resistances in a weighted undirected graph~\cite{spielman2011graph} which is the basic idea behind CAD. This approximation is based on projecting vectors in log-dimensional space with minimal deviations in pairwise distances.
\begin{algorithm}
 \SetAlgoLined
    \SetKwInOut{Input}{Input}
    \SetKwInOut{Output}{Output}
    \Input{$G, E, \epsilon_{RP}$}
    \Output{ $\epsilon_{RP}$ close approximations, $d_{ij}$, of commute time distances between $(i,j)$ for all $(i,j) \in E$}
    $L = D^{G} - G$ \\
    Compute edge-vertex incidence matrices $B$ and $W$ from $G$ \\
    Initialize: $ C = \{ d_{ij} \leftarrow 0   | (i,j) \in E \}$ \\
    $n \leftarrow $ Number of nodes in $G$ \\ 
    $k_{RP} = \left \lceil \log(n/\epsilon_{RP}) \right \rceil$ \\
    \For{$j\gets 1$ \KwTo $k_{RP}$ }{
    Create random vector $q$ \\
     $y = W^{1/2} B q$ \\
     Solve for $z$ in  $Lz = y$  \\
     $\forall (i,j) \in E$, do: $d_{ij}  \leftarrow d_{ij} + (z_i - z_j)^2 $ \\
    }
    return C
    \caption{Commute time distance}
    \label{commutetime_algo}
\end{algorithm}\normalsize

As discussed in Section \ref{background}, the approximate commute time computation relies on efficiently solving a symmetric diagonally dominant system of linear equations. Line 9 of Algorithm \ref{commutetime_algo} shows the linear system that needs to be solved in this case. The Laplacian $L$ can be a really large matrix depending on the size of the problem, and this equation needs to be solved $k_{RP}$ number of times, which makes efficient computation a necessity for this problem. We first discuss the background on such linear system solvers and then describe our re-factored algorithm designed for the distributed framework.

\subsection{Solving SDD Linear Systems}\label{sdd}
An SDD linear system is of the form 
\begin{align}
Mx=b,
\label{eq:linear-system}
\end{align}
 where $M$ is a Symmetric Diagonally Dominant (SDD) matrix ($M \in \mathbb{R}^{n \times n}$) and $b$ is a $n$ dimensional vector. By SDD definition, in Equation~\ref{eq:linear-system}, $M$ is symmetric and has non-positive off-diagonal elements 
\begin{align}
m_{ii} \geq - \sum_{j=1, j\neq i} ^n m_{ij}, \forall i = 1,2,\cdots n.
\end{align}
The solution of this system $M^{-1}b$ requires matrix inversion, which is an $O(n^3)$ operation and can be a bottleneck for large matrices. To reduce the computation time, several approximation techniques are proposed in the literature. Some approximations are designed for matrices of special forms or with specific properties. For a matrix $A$ with norm less than 1, we can approximate the inverse using a finite chain as in Equation~\ref{eq:inverse-chain} with  a small error $\epsilon$: 
\begin{align}
(I-A)^{-1} &= 1 + A + A^2 + \cdots 
= \prod _{k = 0}^d (I+A^{2^k}) + \epsilon.
\label{eq:inverse-chain}
\end{align}
Using this idea Peng and Spielman proposed an approximate SDD linear system solver \cite{peng2014efficient} and presented some directions for parallel construction of the algorithm. The SDD solver algorithm, based on inverse chain, uses a decomposition of the input matrix  $M = D - A$. Here  $D$ is a diagonal matrix with $d_{ii} =  m_{ii}$ and $A$ contains the off diagonal elements: $a_{ij} = - m_{ij}$ for $i\neq j$ and  $a_{ii} = 0$.  The basic inverse chain based solver takes inputs $D$, $A$, $b$, $d$, and computes an approximation, $\hat{x}$, of $M^{-1}b$. The length of the inverse chain, $d$, determines the approximation error which can be set to an arbitrary small number. Using a preconditioner the crude solution can be refined iteratively to obtain an almost exact solution. One such preconditioner is the Richardson preconditioning algorithm \cite{tutunov2015fast} that can reduce the approximation error to a very small number $\epsilon$. However, this algorithm requires $2d$ matrix multiplications which is computationally intensive for large $n$.

Algorithm \ref{alg:sddsolver} describes the SDD solver that we developed for computing the commute time embedding in \name. It consists of two functions: the ChainProduct function that precomputes the chain product $P$ and the EstimateSolution function that uses the precomputed output from ChainProduct in every Richardson iteration. In this refactored SDD solver, the iterations require only matrix-vector multiplications and vector additions thereby significantly reducing the runtime for large matrices.


\begin{algorithm}
 \SetAlgoLined
    \SetKwInOut{Input}{Input}
    \SetKwInOut{Output}{Output}
    \Input{$M, b, d, \delta$}
    \Output{ $\delta$ close approximation, $x*$, of $M^{-1}b$}
    $\bar{\mathbf{P}}_1, \bar{\mathbf{P}}_2 = $ \emph{ChainProduct} $(M, d)$ \\
    return \emph{EstimateSolution} $(\bar{\mathbf{P}}_1, \bar{\mathbf{P}}_2, b, \delta)$ \\
 \vspace{0.5em}
\Fn{ChainProduct ($M, d$)}{
    Construct $D$ and $A$ by from $M$ according to Section \ref{sdd}\\
    $L = D - A$ \\    
	$S = D^{-1/2} A D^{-1/2}$ \\
    $\mathbf{P} = (I+S)(I+S^2)\cdots(I+S^{2^{d-1}})\mathbf{C}$ \\
	$\bar{\mathbf{P}}_1, \bar{\mathbf{P}}_2 = D^{-1/2} \mathbf{P}, D^{-1/2} \mathbf{P} L $  \\      
    return $\bar{\mathbf{P}}_1, \bar{\mathbf{P}}_2$ \\
}
\vspace{0.5em}
\Fn{EstimateSolution $(\bar{\mathbf{P}}_1, \bar{\mathbf{P}}_2, b, \delta)$}{
    $q = \left \lceil \log(1/\delta) \right \rceil$ \\
    $\chi = \bar{\mathbf{P}}_1 b $ \\
    // Preconditioned Richardson iterative scheme\\
    $y_1 = 0$ \\
	\For{$k\gets 1$ \KwTo $q-1$ }{
      $y_{k+1} = y_k - \bar{\mathbf{P}}_2 y_k + \chi$ \\
    }
    return $y_q$
}
    \caption{SDD Solver}
    \label{alg:sddsolver}
\end{algorithm}\normalsize

\subsection{Distributed commute time computation}

Given our focus on large matrix and commute time distance computations for the purpose of large scale graph mining for anomaly detection, we have looked into various software frameworks, namely, MapReduce \cite{dean08mapreduce}, Hadoop,\footnote{http://hadoop.apache.org} and Spark\footnote{http://spark.apache.org}. Due to the iterative nature of the commute time computation, we have chosen Spark for developing the \name framework. Spark's key innovation is the resilient distributed data set (RDD). An RDD is cached, in memory, across computers and supports MapReduce-like parallel computations \cite{zaharia10spark}. Spark manages memory and cores; several distributed operations are lazy. We use Spark to perform distributed matrix multiplication $C = A \cdot B,$ where $A$, $B$, and $C$ are \emph{block} matrices.  A block matrix is a Spark RDD: ((row\_id, col\_id), M).  It should be noted that computing each product block in $C$ requires multiple blocks from each of $A$ and $B$. In Spark's block matrix multiplication supported by the \emph{BlockMatrix}
class ~\cite{bosagh2016matrix}, there is a heavy mixing of blocks between $A$ and $B$ when computing $C$.  Spark's BlockMatrix-multiply attempts complete parallel computation, thus many copies of $A$ and $B$ are created.  This creates a huge shuffle of data, which is extremely memory and disk intensive. In our proposed matrix-multiplication approach, we avoid shuffling of data across workers  by writing $A$ and $B$ to a distributed file system, and computing each block of $C$ by reading all the necessary blocks. This optimized block matrix multiplication not only provides a huge performance boost for this algorithm, but can be useful in any big data application where large matrix multiplication is a bottleneck.

All matrices in Algorithm~\ref{alg:sddsolver} are of the same size as the adjacency matrix, which is $n \times n$. However, in order to scale the SDD solver to very large graphs, we use distributed representation and storage for all matrices as described above. Therefore, we split each $n \times n$ matrix in Algorithm ~\ref{alg:sddsolver} into $\beta$ blocks and store it as an RDD for distributed operations in Spark as shown in Equation~\ref{eq:blockmatrix-rdd}. 
\begin{align}
M \Rightarrow [((1,1), M_{1,1}), ((1,2), M_{1,2}), \cdots, ((\beta,\beta), M_{\beta,\beta}) ]
\label{eq:blockmatrix-rdd}
\end{align}
The two functions in Algorithm~\ref{alg:sddsolver} only use matrix-matrix multiplication or matrix-vector multiplication on the distributed matrices. We compute the $D$ matrix as $D = A \mathbf{1}$ where $\mathbf{1} \in \mathbf{R}^n$ is a column vector. However, the native BlockMatrix multiply operation in Spark requires $O(n^3/p)$ space in the shuffle operation, where $p$ ($=n/\beta$) is the size of each square block. This can be significant for very large matrices (for instance, when $n = 100,000$, a matrix of size 80 GB requires $O(8\text{ TB})$ space with blocks of size 1000. To circumvent this I/O issue, we have developed a shuffle-free implementation of distributed block matrix multiplication using Spark operations. Our method requires $O(n^2)$ space and relies on fast read operations from distributed file system.

The product $C$ of two block matrices $A$ and $B$ of same size can be defined as Equation~\ref{eq:block-matrix-product}, where $A_{i,k} B_{k,j}$ refers to regular matrix multiplication. When $A$ and $B$ are stored in different worker's memory in a distributed network, constructing each product block requires large data shuffle between workers.
\begin{align}
C_{i,j} = \sum_{k=1}^{\beta} A_{i,k} B_{k,j}
\label{eq:block-matrix-product}
\end{align}
Instead of direct data-shuffling between workers, we implement information transfer via a shared distributed storage. Our multiplication procedure involves three steps. 
\begin{enumerate}
\item 
First, all executors write the blocks of $A$ and $B$ to separate files (with names as block ids) in the shared storage. Then the blocks of $C$ are constructed using a MAP operation on the rdd of all block-ids $\{(i,j): 1\leq i \leq \beta, 1\leq j \leq \beta\}$.
\item
The MAP method for $(i,j)$ reads all necessary blocks from $A$ and $B$: $\{A_{i,k}: 1\leq i \leq \beta\}$ and $\{B_{k, j}: 1\leq j \leq \beta\}$.
\item
The MAP method for $(i,j)$ constructs $C_{i,j}$ by Equation~\ref{eq:block-matrix-product}.
\end{enumerate}

\begin{figure}[!ht]
\begin{center}
\includegraphics[width=0.9\textwidth, keepaspectratio]{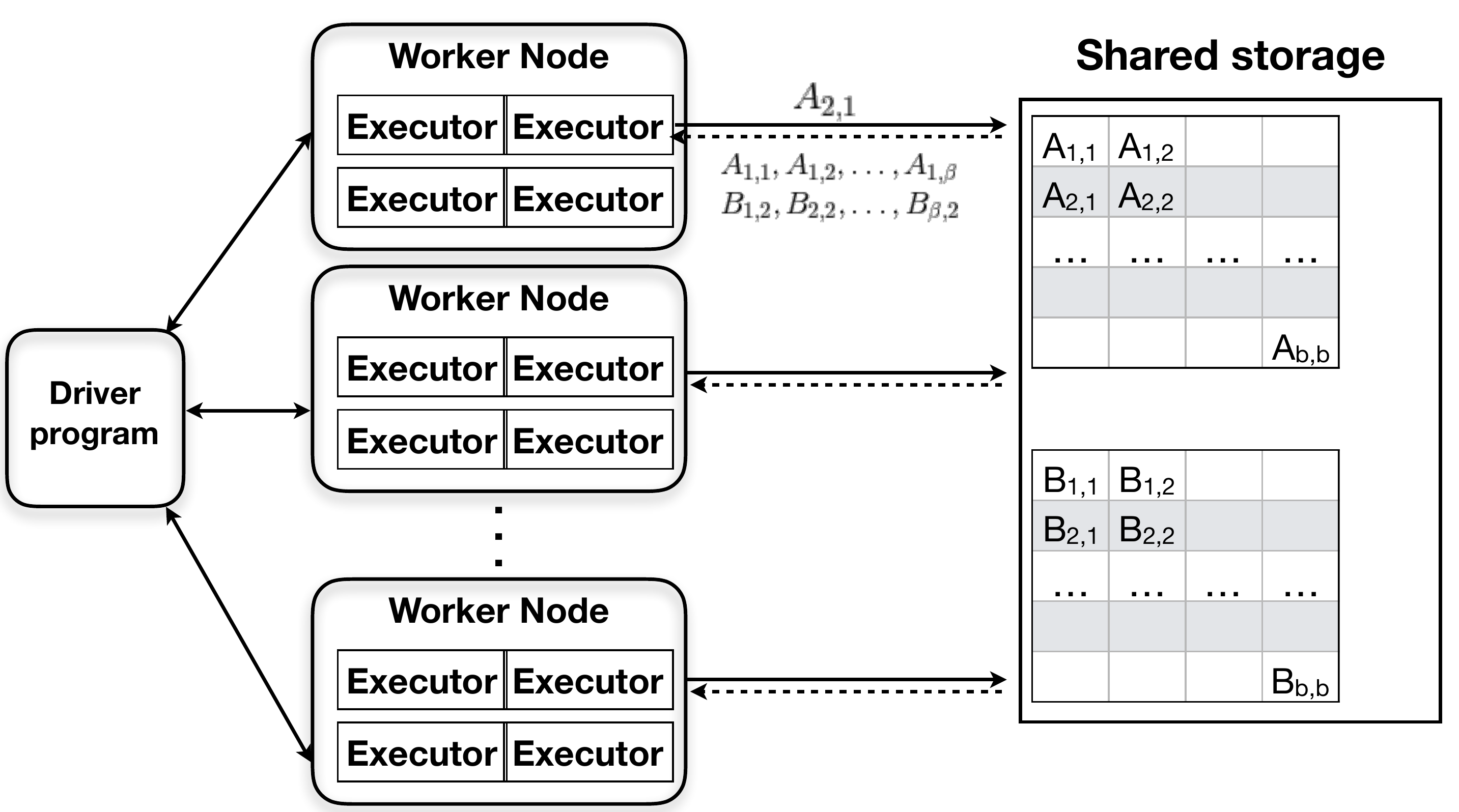}
\end{center}
\caption{Modified distributed matrix multiplication using Spark. Unlike Spark's BlockMatrix multiplication that replicates each block several times for reading by different worker nodes, in our implementation each executor writes one block only once, but reads several blocks from different locations for computing block product.}
\label{fig:matrix-multiplication-nas}
\end{figure}

The overall system architecture is described in Figure \ref{fig:matrix-multiplication-nas}

\begin{algorithm}
 \SetAlgoLined
    \SetKwInOut{Input}{Input}
    \SetKwInOut{Output}{Output}
    \Input{$G, \epsilon_{RP}, \delta$}
    \Output{ Embedding $Z$ for $\epsilon_{RP}$ close approximations, $d_{ij}$}
    $n \leftarrow $ Number of nodes in $G$ \\ 
    $A \leftarrow $ The adjacency matrix of $G$ \\
    $k_{RP} = \left \lceil \log(n/\epsilon_{RP}) \right \rceil$ \\
    Compute $B, W $ according to Equation~\ref{eq:def-B}\\
    Compute Laplacian $L = D - A$\\
    Call Algorithm~\ref{alg:sddsolver}: $\bar{\mathbf{P}}_1, \bar{\mathbf{P}}_2 = $ \emph{ChainProduct} $(L, d)$ \\    
    
    Initialize $Z \gets \mathbf{0} \in \mathbb{R} ^{n \times k_{RP}} $ \\
    \For{$j\gets 1$ \KwTo $k_{RP}$ }{
    Create random vector $q$ \\
     $y = W^{1/2} B q$ \\
     Call Algorithm~\ref{alg:complete-sdd-solve}: $z = $ EstimateSolution $(\bar{\mathbf{P}}_1, \bar{\mathbf{P}}_2, y, \delta)$ \\
     $Z_j \gets z$ // as a column vector \\
     }
    return $Z$
    \caption{CommuteTimeEmbedding}
    \label{alg:commute-time-dist-refactored}
\end{algorithm}\normalsize



Using the efficient SDD solver in Algorithm~\ref{alg:commute-time-dist-refactored}, we  approximately compute the commute time distances between a set of node pairs using the Spielman-Srivastava technique for computing effective resistances~\cite{spielman2011graph}. For each random vector $q$ (line 9 in Algorithm \ref{alg:commute-time-dist-refactored}), SDD system $Lz = y$ is solved using the graph Laplacian $L$ and a vector $y$. Along with the chain product $\mathbf{P}$, we precompute two other intermediate results $\bar{\mathbf{P}}_1$ and $\bar{\mathbf{P}}_2$ such that the loop over random vectors only require matrix-vector multiplications, which is done using our updated block matrix multiplication. 




 



\subsection{Anomaly Detection using CADDeLaG}

Algorithm \ref{alg:caddelag} describes the complete commute time based anomaly detection framework \name~\footnote{\url{https://github.com/olemengshoel/caddelag}}. We introduce a few modifications to the CAD algorithm to obtain an efficient implementation. CAD attempts to find a minimal set of edges, such that the total change in the distance metric ($\Delta E_t$) of the remaining edges is lower than a user-defined threshold $\delta$. As the distance metric is based on commute time distances ($\Delta d_t$), it computes the pair-wise commute time distances for all pairs in the graph. However, computing commute time distances for all pairs of nodes is $O(n^2)$ operation. Since,  $\Delta E_t = |\Delta A_t| |\Delta d_t|$ where $\Delta A_t$ is the change in edge weights of the given node pair, we only need to compute $\Delta d_t $ for which $\Delta A_t \neq 0$. In summary, \name first identifies the node pairs with non-zero change in weights and computes commute time distances for only those pairs. Next, \name computes  $\Delta E_t$ for all those nodes and finds the anomalous edges according to the optimization formulation of CAD.

\begin{algorithm}
 \SetAlgoLined
    \SetKwInOut{Input}{Input}
    \SetKwInOut{Output}{Output}
    \Input{$G_1, G_2$ \qquad \qquad \textbf{Accuracy parameters}: $\epsilon_{RP}, \delta$}
    \Output{Anomalies in $G_1 \rightarrow G_2$}
    Call Algorithm~\ref{alg:commute-time-dist-refactored}: $Z_1$ = CommuteTimeEmbedding($G_1, \epsilon_{RP}, \delta$) \\
    Call Algorithm~\ref{alg:commute-time-dist-refactored}: $Z_2$ = CommuteTimeEmbedding($G_2, \epsilon_{RP}, \delta$) \\
    Compute commute-time distance matrices $D_1, D_2$ from $Z_1, Z_2$ \\
    $A_1,A_2:$ adjacency matrices representing $G_1$ and $G_2$\\
    $\Delta E = |A_1 - A_2| \odot |D_1 - D_2|$, where $\odot$ is Hadamard product \\
    Node anomaly scores $F_i = \sum_{j=0}^{n-1} \Delta E_{i,j} $ \\
    Return the set of top anomalies based on high $F_i$ values.
    \caption{\name}
    \label{alg:caddelag}
\end{algorithm}\normalsize

\section{Performance of CADD\lowercase{e}L\lowercase{a}G}
\label{expmts}

In this section we describe the runtime and storage complexity of \name. Then we present experimental analysis of performance, to illustrate the effect of various user-defined parameters.

\subsection{Complexity analysis}\label{complexity}
Here we analyze the runtime complexity of \name in the Spark computation framework. We denote the total number of parallel processing units or Spark executors as $S$. Spark allocates parallel MAP tasks to all executors and supports several reduction operations. Moreover, certain steps of \name (e.g. matrix multiplication) includes writing or reading the distributed block matrices from shared storage. We denote the runtime of $x$ parallel read-write operations in the distributed file system by $\phi(x)$. 
In the high performance distributed network used in our experiments, the read-write time is much smaller compared to operations on individual blocks. More details about our computing infrastructure is provided in Section~\ref{sec:experimental-analysis}. 

First we analyze the complexity of Algorithm~\ref{alg:commute-time-dist-refactored} which is the major time-intensive part of \name. Computing $B, W$ involves random projection of the adjacency matrix $A$ and summing the columns of the projected matrix. We implement this step as a Map and a ReduceByKey operation. The runtime of this step is $O(p^2\beta^2/S$ $+ \phi(\beta^2))$.	We also implement multiplication of a distributed matrix with a vector by one Map followed by a ReduceByKey operation. Therefore computing $D=A\mathbf{1}$ takes $O(p^2\beta^2/S$ $+ \phi(\beta^2))$ time. Addition or subtraction operations on $D$ and $A$ takes $O(p^2\beta^2/S)$ time. The total complexity of lines 1-5 is $O(p^2\beta^2/S + \phi(\beta^2))$. 

The second part of of Algorithm~\ref{alg:commute-time-dist-refactored} is computing the product of matrix-chain by Algorithm~\ref{alg:sddsolver}. Our distributed matrix multiplication procedure includes writing all blocks to shared storage in $\phi(\beta^2)$ time, and constructing each product block independently in parallel. The Map task of one block construction consists of reading $2\beta$ blocks and computing their inner product (see Equation~\ref{eq:block-matrix-product}). Multiplication of individual matrix-blocks require $O(p^{2+\zeta})$ time (where $0 < \zeta < 1$), and therefore the complexity of one block construction is $O(\beta p^{2+\zeta}  + \phi(2\beta))$. Since executor needs to construct $\beta^2/S$ blocks, the total complexity of distributed matrix multiplication is $O(\beta^3 p^{2+\zeta}/S  + \phi(2\beta))$. These multiplications are repeated $d$ time to compute the chain-product. The overall runtime of \emph{ChainProduct} is $O(d\beta^3 p^{2+\zeta}/S  + \phi(2\beta))$.

The last part of of Algorithm~\ref{alg:commute-time-dist-refactored} makes $k_{RP}$ calls to \emph{EstimateSolution} which runs a loop for Richardson iterative update. The independence of the $k_{RP}$ random vectors allows us to compute columns of $Z$ in parallel and thus, we can reduce the two loops into a single loop of Richardson iterative update where $y_k$ and $\chi$ are matrices  of size $n \times k_{RP}$. The runtime of lines 7 - 14 of Algorithm~\ref{alg:commute-time-dist-refactored} reduces to $O(q p^2\beta^2/S + \phi(\beta^2))$. 

For \name (Algorithm~\ref{alg:caddelag}), we recognize that the matrices in lines 4-5 can be constructed  independently at the block level. Using Map operations we can compute $\Delta E$ from $G_1, G_2, D_1, \text{and} D_2$ in $O(p^2\beta^2/E + \phi(\beta^2))$ time.

The total runtime of   Algorithm~\ref{caddelag}, is  $O(d\beta^3 p^{2+\zeta}/E  + q p^2\beta^2/E + \phi(\beta^2))$ or $O(d\beta^3 p^{2+\zeta}/E + \phi(\beta^2))$ since $d\beta^3  > q \beta^2$. Clearly, $E$ or the size of the Spark cluster has a big impact on the runtime. For a large graph \name can take a few months with small $E$ and can also complete within a few hours with sufficiently large $E$. Moreover, the block size $p$ also affects the runtime significantly because it affects the number of blocks $\beta^2$ and the number of read-write operations. We choose $E$ to be proportional to the number of blocks $\beta^2$, and $p = \Theta(n^{0.5})$. Therefore, we get $ E = \beta^2 / M, \beta = \Theta(n^{0.5})$ and thus, $E = \Theta(n/M)$. Substituting these values, we get the overall runtime of \name to be $O(dn^{1.5 + \zeta})$. The read-write time is much less compared to this time and thus ignored. Efficient implementation of matrix multiplication, as present in \emph{numpy} library can potentially reduce $\zeta$ below 0.5 and the literature presents various methods of matrix multiplication in $O(n^{2.5 - x})$ time, where $0 < x < 0.5$~\cite{coppersmith1987matrix,le2014powers}. 

\subsection{Experimental analysis}
\label{sec:experimental-analysis}
We use NASA's high performance computing infrastructure for all our experimental needs. Specifically, we used the Pleiades\footnote{\url{https://www.nas.nasa.gov/hecc/resources/pleiades.html}} supercomputer. It contains the following  Intel Xeon processors: E5-2680v4 (Broadwell), E5-2680v3 (Haswell), E5-2680v2 (Ivy Bridge), and E5-2670 (Sandy Bridge). We use Broadwell nodes with 28 cores and 128GB of memory. These machines access a shared storage with a total of 40 petabytes (PB) of disk space, serving thousands of cores, under the Lustre-based filesystems managed under Lustre software version 2.x.\footnote{See \url{http:/lustre.org/}.}  The nodes are connected to the cluster file system mass storage device via InfiniBand with a bandwidth of 56 Gb/sec. Pleiades utilizes a PBS scheduler for distributed job scheduling through the cluster manager. All nodes run the  SGI ProPack operating systems for Linux kernel version 3.2. For all distributed experiments we use Apache Spark version 2.1 with Hadoop Yarn 2.6 as the cluster manager. We use the Oracle Java runtime, Java SE 8u and run Spark in the cluster mode. The Spark driver runs on one dedicated coordinating node with only one Spark executor. The workers, however, run multiple Spark executors. 

We modify a few default Spark configurations to cater to the high memory requirement of the matrix multiplication step. We allocate 8GB memory and 2 cores for each Spark executor and maximize the number executors per host. Moreover, the matrix multiplication requires high parallel-read write from/to the distributed storage, managed by Lustre. We tune ``stripe count'' to optimize parallel read-write performance. 

\begin{figure}
\centering
\begin{subfigure}{0.5\textwidth}
\centering
\includegraphics[width=0.95\textwidth, keepaspectratio]{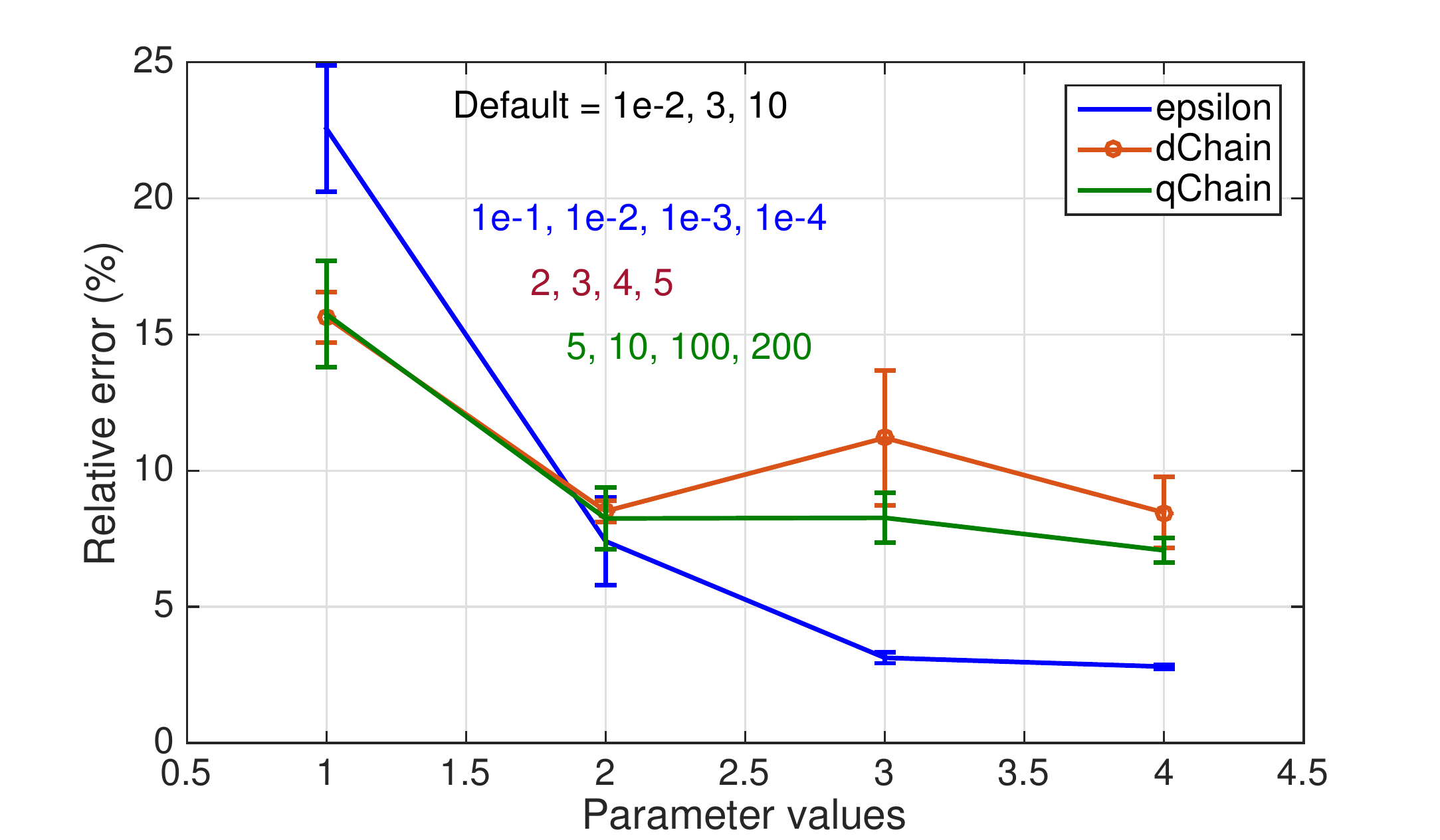}
\caption{ With high $\epsilon_{RP}$, relative\\ error does not go below 6\% for\\any value of dChain and qChain.}
\label{fig:error-vs-parameters-CHAIN-e2}
\end{subfigure}%
\begin{subfigure}{0.5\textwidth}
\centering
\includegraphics[width=0.95\textwidth, keepaspectratio]{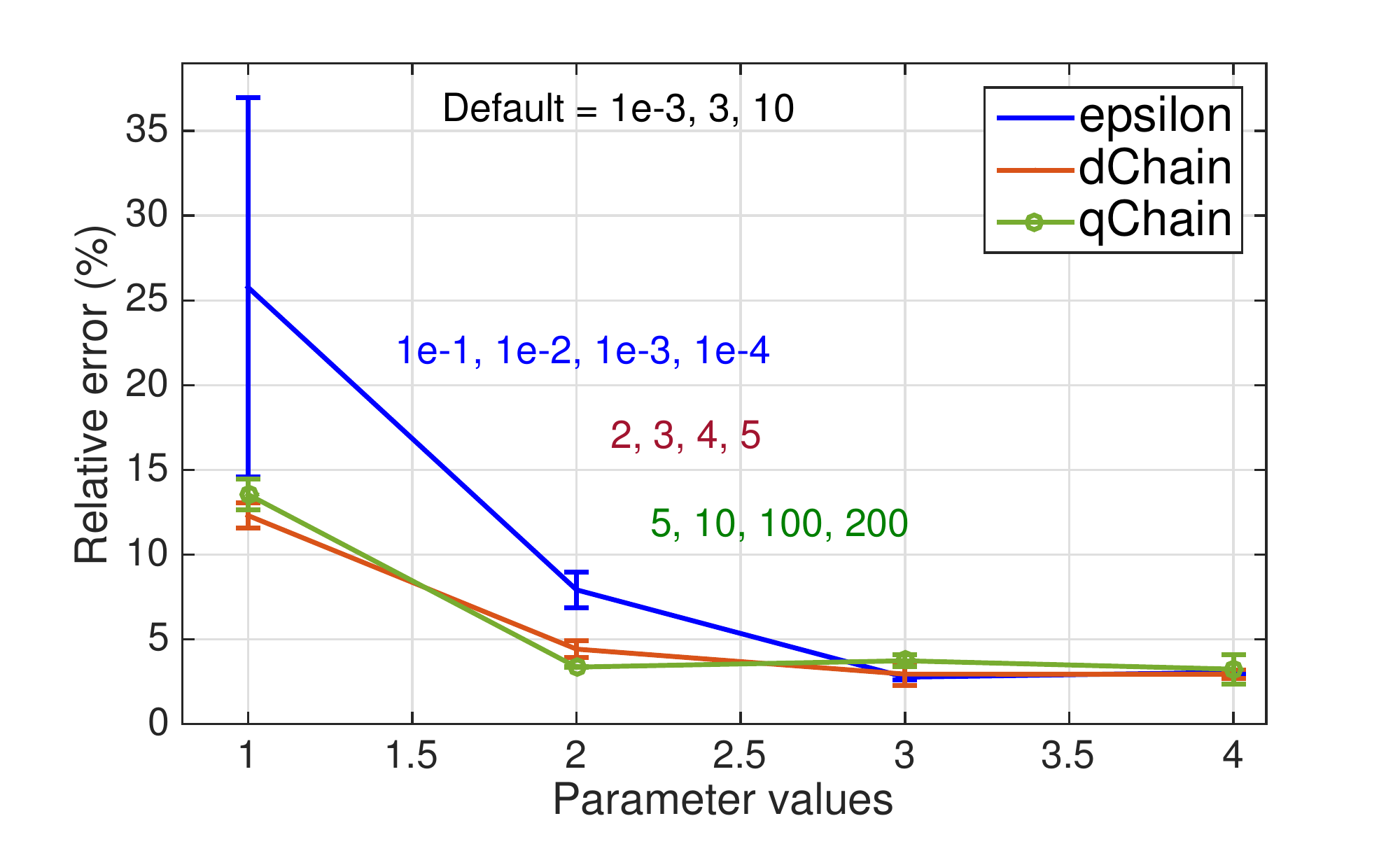}
\caption{ For $\epsilon_{RP}=1\mathrm{e}{-3}$, the relative error decreases substantially, even for less strict values of $d$ and $q$. }
\label{fig:error-vs-parameters-CHAIN-e3}
\end{subfigure}
\caption{Effect of epsilon ($\epsilon_{RP}$), dChain ($d$), and qChain ($q$) on the accuracy of \name. While varying one parameter, we keep the rest at ``default'' settings for each Figure.}
\end{figure}

\subsubsection{Data description}\label{data}
For experimental performance analysis, \name is tested on several synthetic and real world data sets. In our analysis we are interested in not only the scalability of \name, but also the accuracy in terms of its ability to localize anomalous edges and nodes responsible for significant changes in graph structure with respect to the centralized CAD algorithm.
Since the real data sets only provide anecdotal evidence of anomalies, we create a synthetic data set from Gaussian mixtures to quantitatively analyze the performance of \name. The three data sets on which we run \name are (i) synthetic data set created from a mixture of Gaussians, (ii) the 2016 US election donation data set, and (iii) a worldwide precipitation data set. We also ran experiments on the Stackoverflow\footnote{ \url{https://stackoverflow.com/}} social network which is a sparse graph (unlike the other data sets). We do not include the results for Stackoverflow anomaly analysis here due to lack to space. 
\\
\\\noindent\textbf{Synthetic Data}\label{synthetic} We follow the method described in \cite{sricharan2014localizing} to draw random samples from a 2-dimensional Gaussian mixture distribution with 4 components and compute the Euclidean distance for all pairs of points in this distribution. We construct a matrix $P$ where $P(i,j) = \exp(-d(i,j))$.  The graph corresponding to this matrix  contains nodes belonging to 4 different clusters, with strong intra-cluster edges and weaker inter-cluster edges. We perturb this matrix $P$ by adding a small amount of random noise to the data, and compute the matrix $Q$ in an identical manner to $P$. We also construct a random matrix $R$, where each entry $R(i,j)$ is given by
\begin{eqnarray}
R(i,j) = \begin{cases} 0 & \mbox{with probability } p=0.95  \\ \mbox{$u(i,j)$} & \mbox{with probability } p=0.05, \end{cases}
\nonumber
\end{eqnarray}
where $u(i,j)$ is a random number drawn uniformly between $0$ and $1$. We then consider a dynamic graph sequence with two temporal instances $\{A_t, t=1,2\}$, with $A_1 = P$ and $A_2 = Q+(R+R')/2$. In the transition from $A_1$ to $A_2$, we keep track of all edges for which $R(i,j) \neq 0$ such that $i,j$ belong to different clusters. We consider these edges and the associated nodes to be anomalous because these edges establish ties between nodes belonging to different clusters, thereby contributing to anomalous change in graph structure. This data set is generated at various sizes for our experiments. 
\\
\\
\noindent\textbf{Election Data}
The U.S.~Federal Election Commission (FEC)~\footnote{\url{https://www.fec.gov/}} administers and enforces U.S.~campaign finance laws for campaigns for the U.S.~Congress and Presidential elections. Our experimental data set consists of individuals who contributed to the the 2016 Presidential election cycle. Although most of the contributions are made to committees supporting different candidates, we consolidate the donations with respect to the political affiliations: namely Democrats, Republicans, and Others. To understand the transition of public support (sentiment) from the early phase of elections till the final election date, we construct two temporal graphs for donations: the first graph represents the early phase Jan 2015 - May 2016 and the second one is for final phase June 2016 - Nov 2016. The nodes in the graph are all individuals who donated in both time periods. We construct graphs using two different distance measures. In the first setting the edges of the undirected graph signify the similarity of support between two individuals, i.e. if two individuals donate to the same party there is an edge between them whose strength is defined by the minimum of the two donation amounts for that phase. In the second setting the donation amounts are divided into three categories: low ($\leq \$1000$), medium ($>\$1000$ and $\leq\$10000$), and high ($>\$1000$) and individuals are connected to each other through edges only if they donate to the same party and belong to the same category. In this setting we also update the weights to be in the logarithmic scale of the donation amounts instead of the actual scale to suppress the effect of donation amounts in the anomaly calculations. The number of nodes  is 555924. The number of edges for the two phases are $2.3\mathrm{E}{11}$ and $2.47\mathrm{E}{11}$ for the first setting, and $1.44\mathrm{E}{11}$ and $1.68\mathrm{E}{11}$	for the second.
\\




\noindent\textbf{Climate Data}
The world-wide precipitation (rainfall) data from NCEP/NCAR\footnote{\url{http://www.cdc.
noaa.gov/data/gridded/data.ncep.reanalysis.html}} consists of monthly averages of observations between 1982 and 2002 recorded at a spatial resolution of $0.5^\circ$ on the entire earth's grid. The data set has 259,200 geolocations with 67,420 locations having recorded precipitation values at any instance. For detecting abnormal global changes in the precipitation pattern, we create a fully connected precipitation graph that has 259,200 nodes and the edges weights are based on the distance function $\exp\left(-{\left\Vert p_i-p_j\right\Vert^2}/{2\sigma^2}\right)$ with $p_i$ and $p_j$ corresponding to precipitation levels recorded at $i$ and $j$ using the optimized kernel bandwidth $\sigma$ value of 388.

\begin{figure*}
\centering
\begin{subfigure}{0.33\textwidth}
\centering
\includegraphics[width=0.95\textwidth, keepaspectratio]{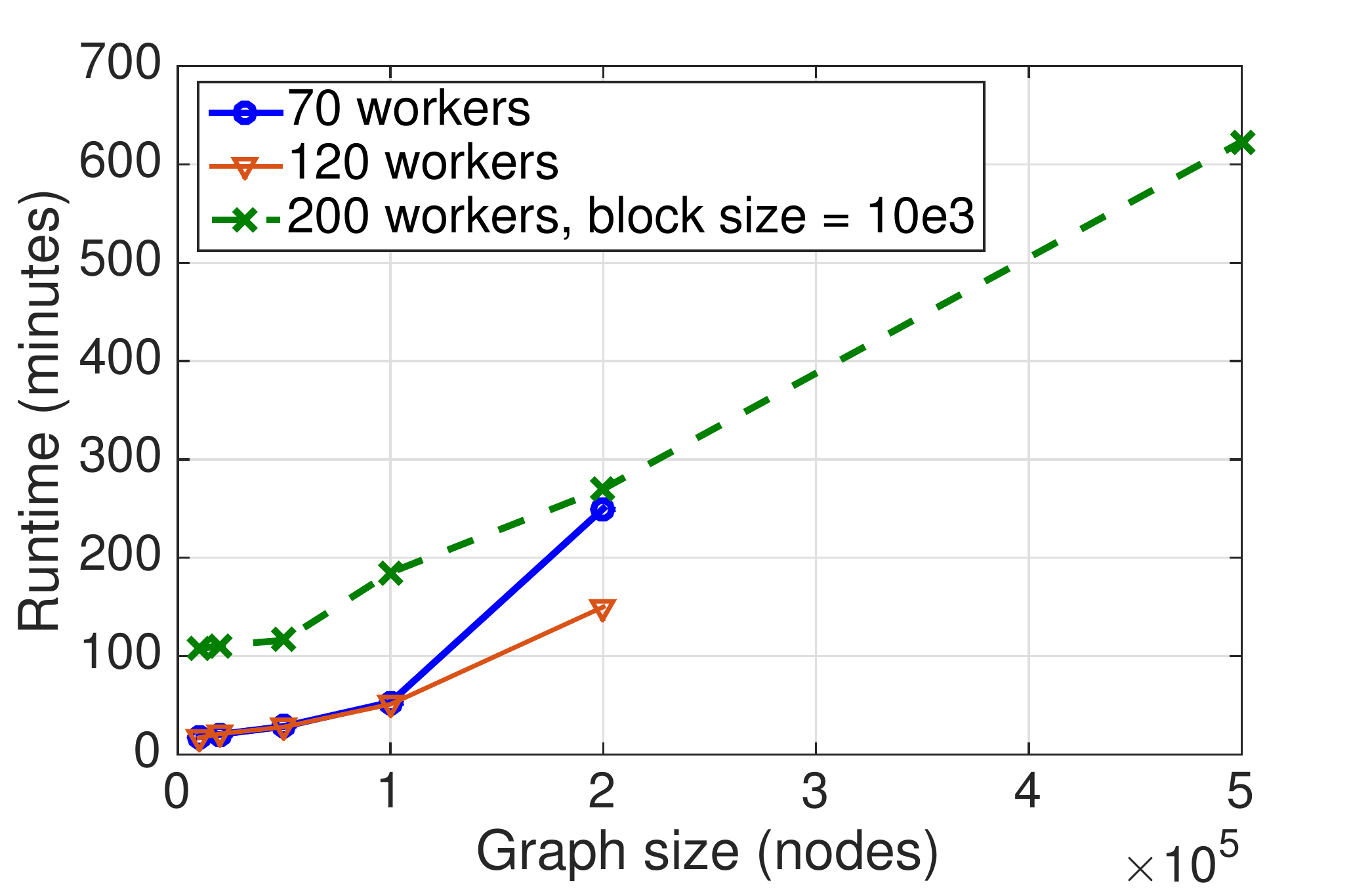}
\caption{Block size is a function of graph size \\and worker nodes for 70 and 120 workers. \\It is fixed for 200 workers due to limited \\resource availability.}
\label{fig:runtime-vs-graph-size}
\end{subfigure}%
\begin{subfigure}{0.33\textwidth}
\centering
\includegraphics[width=0.95\textwidth, keepaspectratio]{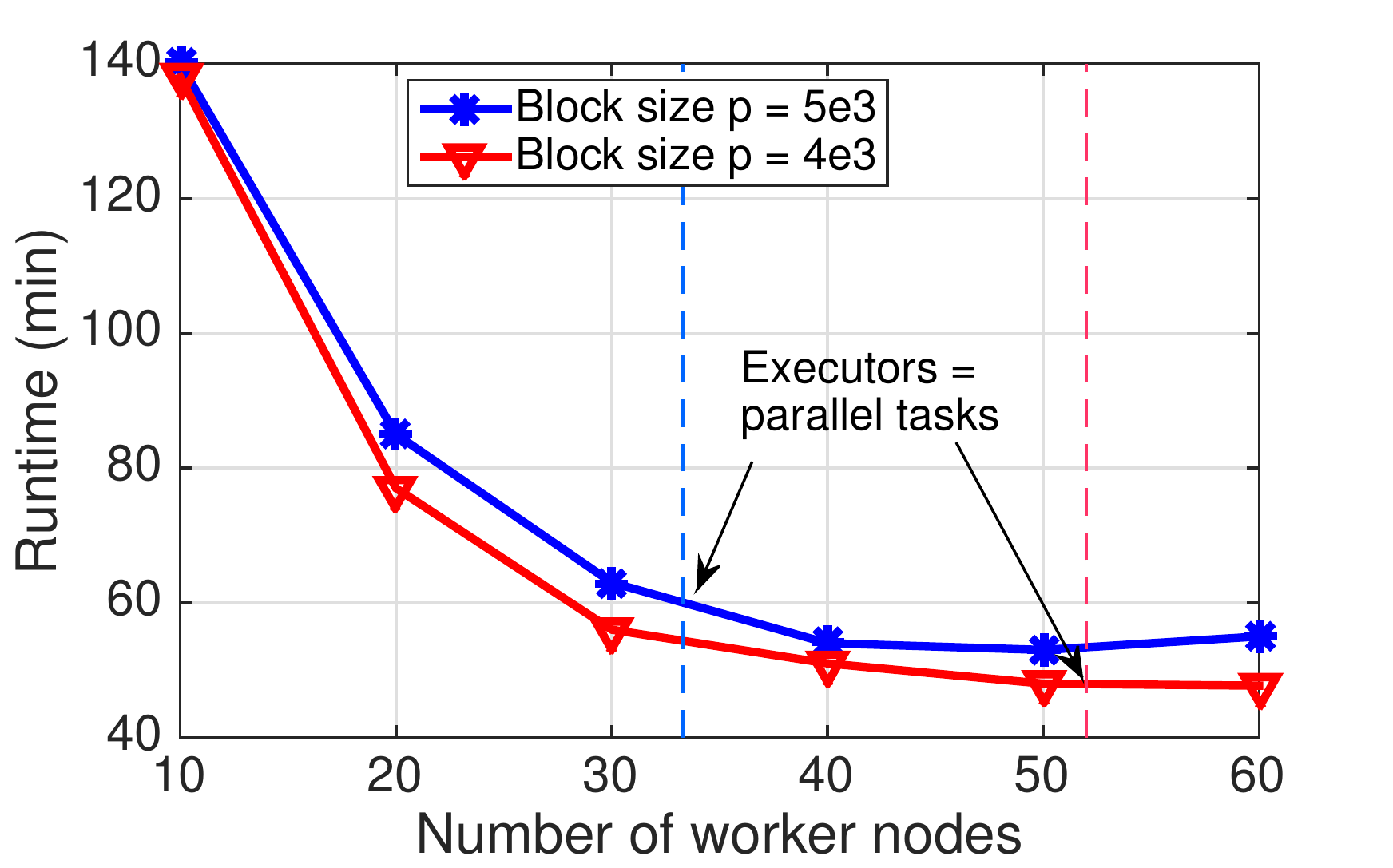}
\caption{The optimal number of worker nodes \\is equal to the number of parallel tasks \\for a given graph} 
\label{fig:runtime-vs-cluster-size}
\end{subfigure}%
\begin{subfigure}{0.33\textwidth}
\centering
\includegraphics[width=0.95\textwidth, height = 0.165\textheight]{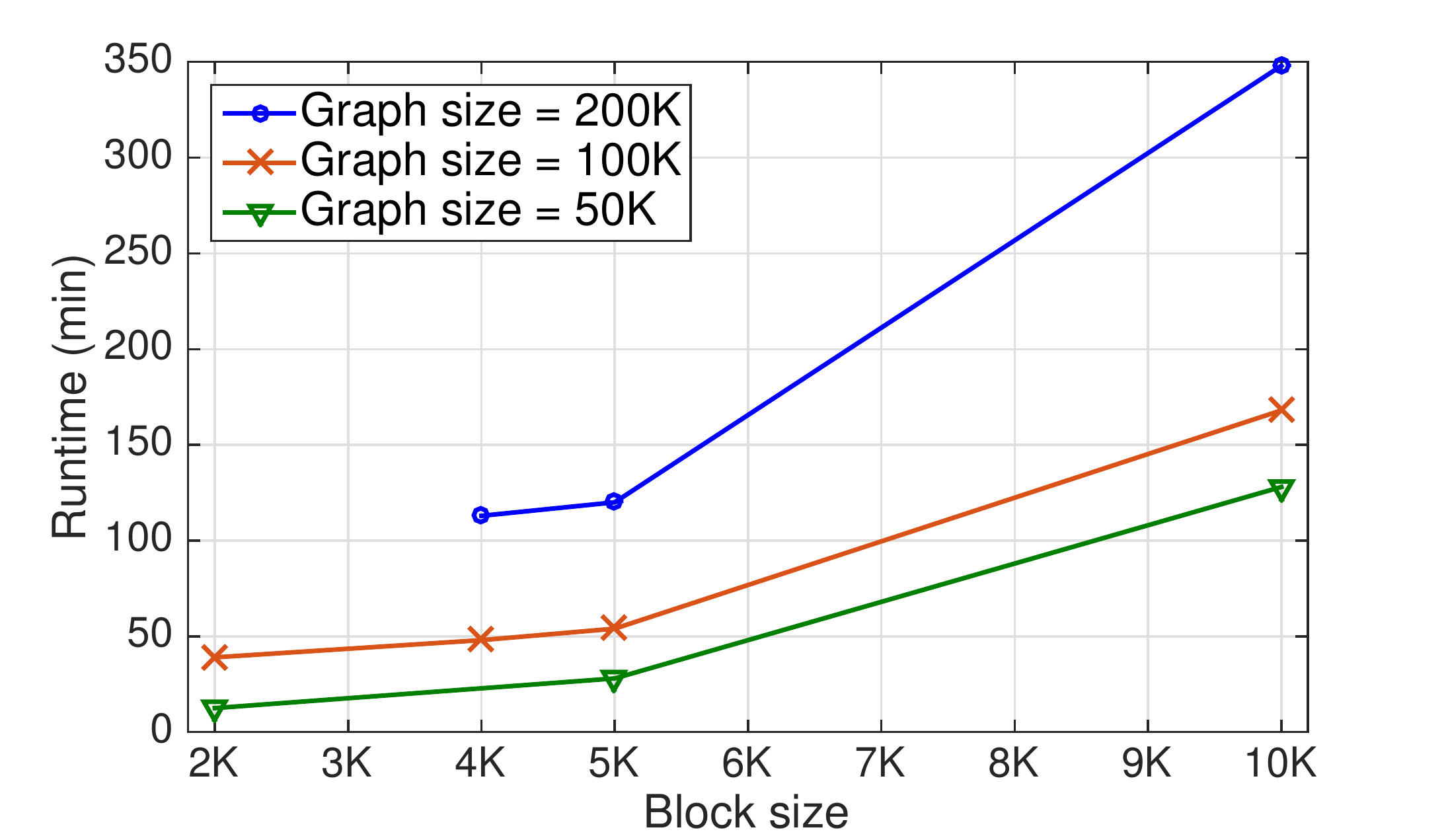}
\caption{Maximum parallelism by matching number of executors to parallel tasks (except for 200K graph size) ensures variability in running time is primarily due to block size variation}
\label{fig:runtime-vs-block-size2}
\end{subfigure}
\caption{Variation in running time of commute time distance computation with respect to (a) problem size, (b) compute cluster size, and (c) block size.}
\end{figure*}


\subsubsection{Accuracy}
The goal of the first set of experiments is to understand how the accuracy of \name depends on various user defined parameters, \emph{viz.} on the length of the chain $d$ in the inverse computation and the number of iterations $q$ in the Richardson preconditioning scheme (Algorithm \ref{alg:sddsolver}), and lastly the allowed error $\epsilon_{RP}$ deciding the embedding dimension of commute time (Algorithm \ref{alg:commute-time-dist-refactored}). Here accuracy measures the correctness of the distributed commute time computation in comparison to direct eigen decomposition of the graph Laplacian and centralized approximate SDD solver based computation \cite{koutis2011solving}. The metric we track is relative error, defined as 
\begin{align*}
\text{Relative error} = \frac{\text{\name error} - \text{Baseline error}}{\text{Baseline error}}
\end{align*}
Since the baseline methods are centralized and do not scale, these results are reported on synthetic data generated as described in Section \ref{synthetic} with $n=2000$. The accuracy results are shown in Figures \ref{fig:error-vs-parameters-CHAIN-e2} and \ref{fig:error-vs-parameters-CHAIN-e3}.



In Figure \ref{fig:error-vs-parameters-CHAIN-e2}, the default choices for $\epsilon_RP$, $d$, and $q$ are $1\mathrm{e}{-2}$, $3$, and $10$ respectively. For each parameter sensitivity analysis, we only alter its value, while keeping the other two fixed at their default values. We notice that the relative error does not go down below 6\% when $\epsilon_{RP}$ is set to $1\mathrm{e}{-2}$. On the other hand, keeping $\epsilon_{RP}$ at $1\mathrm{e}{-3}$, we can even choose slightly inferior values of $d$ and $q$ without losing any significant accuracy in the commute time distance computation, as is evident from Figure \ref{fig:error-vs-parameters-CHAIN-e3}. Therefore, we can conclude that the embedding dimension is the biggest influencing factor of computational accuracy for \name.

\subsubsection{Scalability}
In this section we present the scaling properties of \name with increasing problem size, under various resource constraints and design choices. All experiments are performed on synthetic data graphs generated as described above. In Figure \ref{fig:runtime-vs-graph-size}, we show how the run time increases when the size of the graph (number of nodes, edges) increases, with compute resources remaining constant. Figure \ref{fig:runtime-vs-graph-size} shows the results for three different cluster sizes we experimented with: a 70-worker,  a 120-worker and a 200-worker infrastructure. In the first two cases, we only show results when the problem size increases up to 200,000 nodes. In the 200-worker setting, we have increased the graph size to up to 500,000. It should be noted that since these graphs are dense, a 10-fold increase in the number of nodes indicates a 100-fold increase in the number of edges. Since commute time embedding is computed in the edge space, we are dealing with an order of $1\mathrm{e}{10}$ edges (and therefore, commute time distance computations) in this experiment. It can be observed that the running time increases linearly on an average with a quadratic increase in problem size. The increase in running time is not only a function of the problem size, but also depends heavily upon the choice of the block size (please see Section \ref{data}), which in turn is dictated by the availability of resources. 

In our second scalability experiment we show how the run times vary with availability of resources, given a fixed problem size and for a chosen block size. Figure~\ref{fig:runtime-vs-cluster-size} shows the results for this experiment for a synthetic graph having 100,000 nodes and $1e10$ edges. For each setting of block size we observe that the running time falls exponentially when the number of workers increases initially. The performance gain slows down and ultimately stops after a certain number of workers is reached. The three phases of this performance graph (exponential improvement, slow improvement, no improvement) can be explained as follows. When the number of workers is significantly smaller than the number of blocks (of the graph matrix), there is a wait time for job scheduling and the computation is not fully parallel. As more workers become available, the wait time for job scheduling reduces until it saturates at the point where the number of workers is enough to accommodate as many executors as the number of blocks. This is shown using the dotted red line in Figure~\ref{fig:runtime-vs-cluster-size}. Beyond this point there is not much performance gain in increasing the number of workers.

In our last experiment we demonstrate how the run times vary with the choice of the block size, given a fixed problem size and with constant amount of available resources. Figure~\ref{fig:runtime-vs-block-size2} shows the results for this experiment for dense synthetic graphs having three different sizes. To make sure we observe only the effect of block size, and not the amount of resources, we maximize parallelism by matching the number of executors to number of parallel tasks in each case, except when the input graph size is 200K and the block size is $\leq 4K$, in which case the amount of resources needed is extremely high and is much more than the amount of available resources. The general observation for this experiment is that smaller the block size, the better is the performance. However, block size cannot be made infinitely small since it increases the amount of worker nodes needed, which is limited for all practical purposes. Increasing the block size hinders performance due to longer I/O times. Therefore the optimal block size is usually the smallest that can be set, keeping in mind the compute resource constraints.



\section{Anomaly detection using CADD\lowercase{e}L\lowercase{a}G}\label{anomalies}

In this section we report the results of anomaly detection on the two real-life data sets: the worldwide precipitation data set and the election data set.

\subsection{Precipitation anomalies}

\begin{figure}
\centering
\includegraphics[width=0.9\textwidth,keepaspectratio]{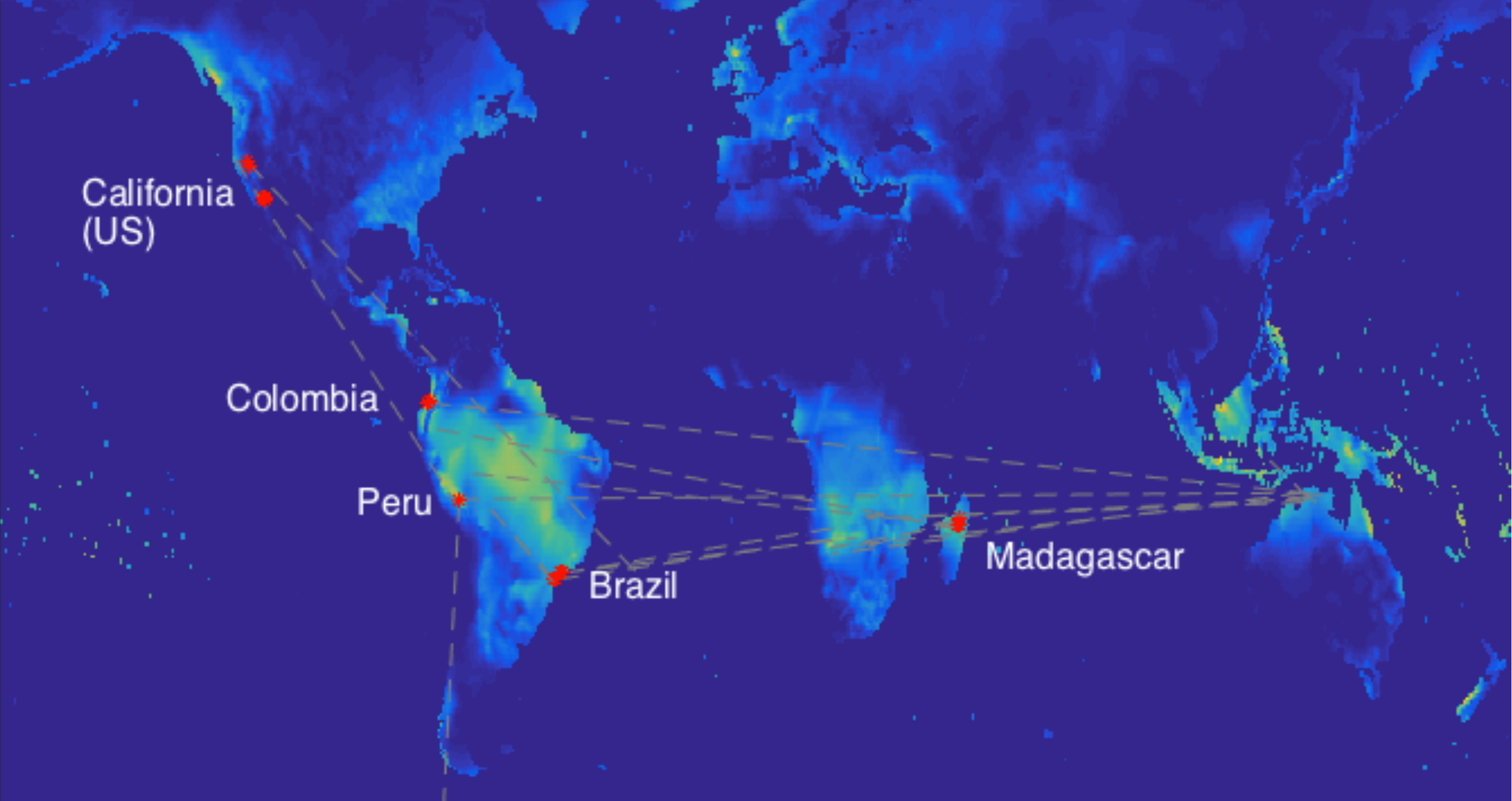}
\caption{Anomalous locations in worldwide precipitation graph for January 1994-95.}\label{fig:anomalies_climate}
\end{figure}

In the precipitation data set, we compare differences in monthly observations over the years. In this paper we report anomaly results for the month of January  for years 1994-1995. We choose this pair of temporal graphs in order to compare the anomalies from \name with those reported in \cite{sricharan2014localizing}. Figure \ref{fig:anomalies_climate} shows the top anomalies found by \name using red stars on the global map. Based on anecdotal evidences all of these locations had major precipitation events in the month of January in 1994 or 1995. The top ranking anomaly is in northern California which suffered an historic flood on January 1995, in which 28 people were killed and property worth USD 1.8 billion was destroyed\footnote{\url{https://pubs.usgs.gov/fs/1995/0062/report.pdf}}. A second location identified by \name is in Madagascar island in Africa where cyclone Geralda hit in 1994\footnote{See 1997 book Mitigating the Millennium: Proceedings of a Seminar on Community. edited by Jane Scobie, pp. $53-55$} and has been touted as one of the worst cyclones hitting that area. Locations  found in South Pacific such as Colombia and Peru and Brazil are all part of the 1994-95 El Ni\~{n}o phenomenon discussed in \cite{sricharan2014localizing}. The edges going out of each of these anomalous locations point to the biggest changes in mutual relationships among the points. Northern Territory in Australia had a relatively normal rainfall season in 1994-95. Therefore, its relationship with both Africa and South American countries such as Colombia, Peru, and Brazil changed due to  El Ni\~{n}o affecting the South American countries and the historic floods in Africa in 1994. Similarly, California's relationship changed with the more stable Southern part of the globe due to anomalous rainfall in 1995. These results show the importance of \name over the existing CAD algorithm. Since CAD cannot handle fully connected graphs, only 10-nearest neighbor graphs were analyzed for anomalies in \cite{sricharan2014localizing}. Such sparsification leads to loss of meaningful information. In this case, although CAD found the El Ni\~{n}o phenomenon, one of the worst California floods got left out from the results since the graphs analyzed by CAD were sparsified to keep only the 10-nearest neighbors of every node in the graph. 

\subsection{Election anomalies}

\begin{figure}
\centering
\begin{subfigure}{0.3\textwidth}
\centering
\includegraphics[width=1\textwidth, height = 0.09\textheight]{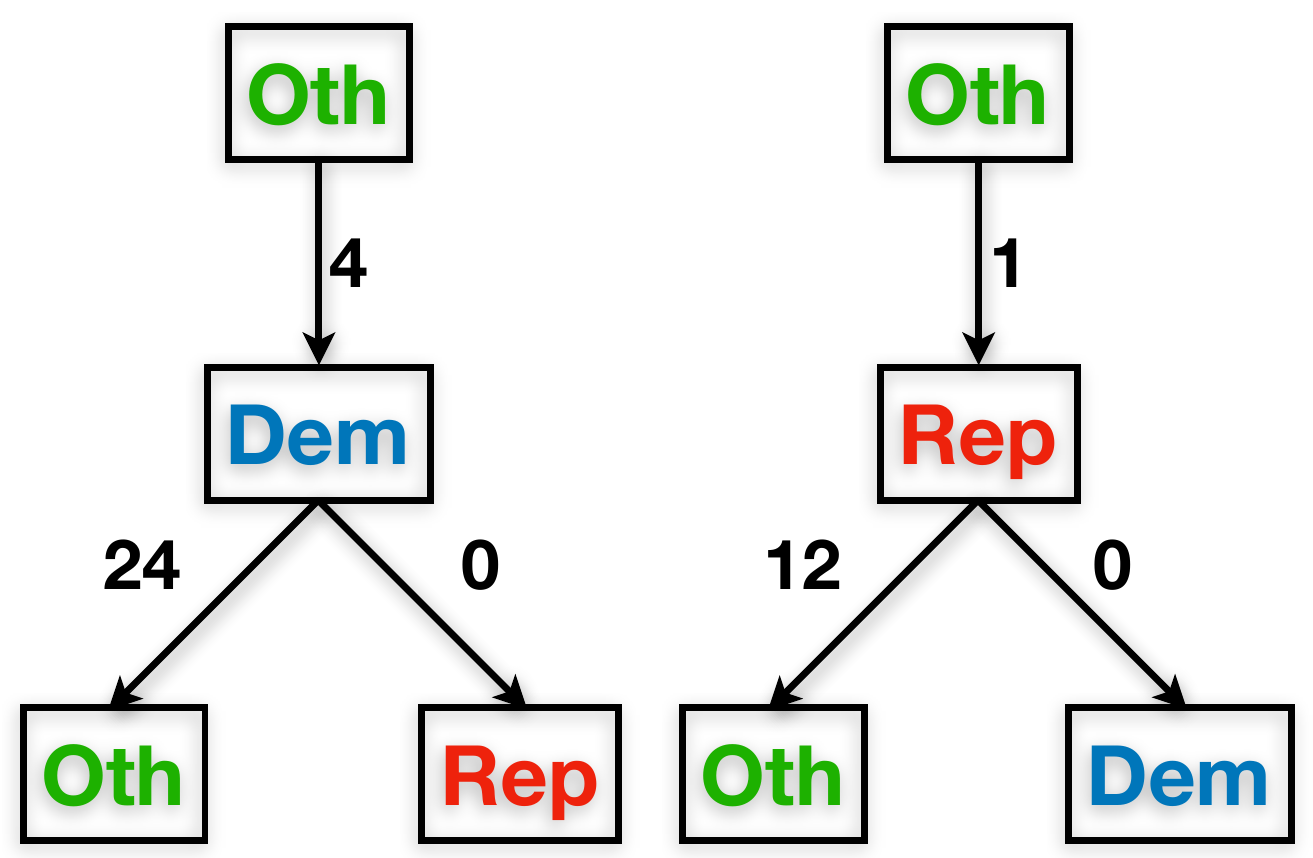}
\caption{Democrats lost more donors than Republicans among the top 100 anomalies}\label{fig:election_trends_all}
\end{subfigure}\hfill
\begin{subfigure}{0.33\textwidth}
\centering
\includegraphics[width=0.8\textwidth]{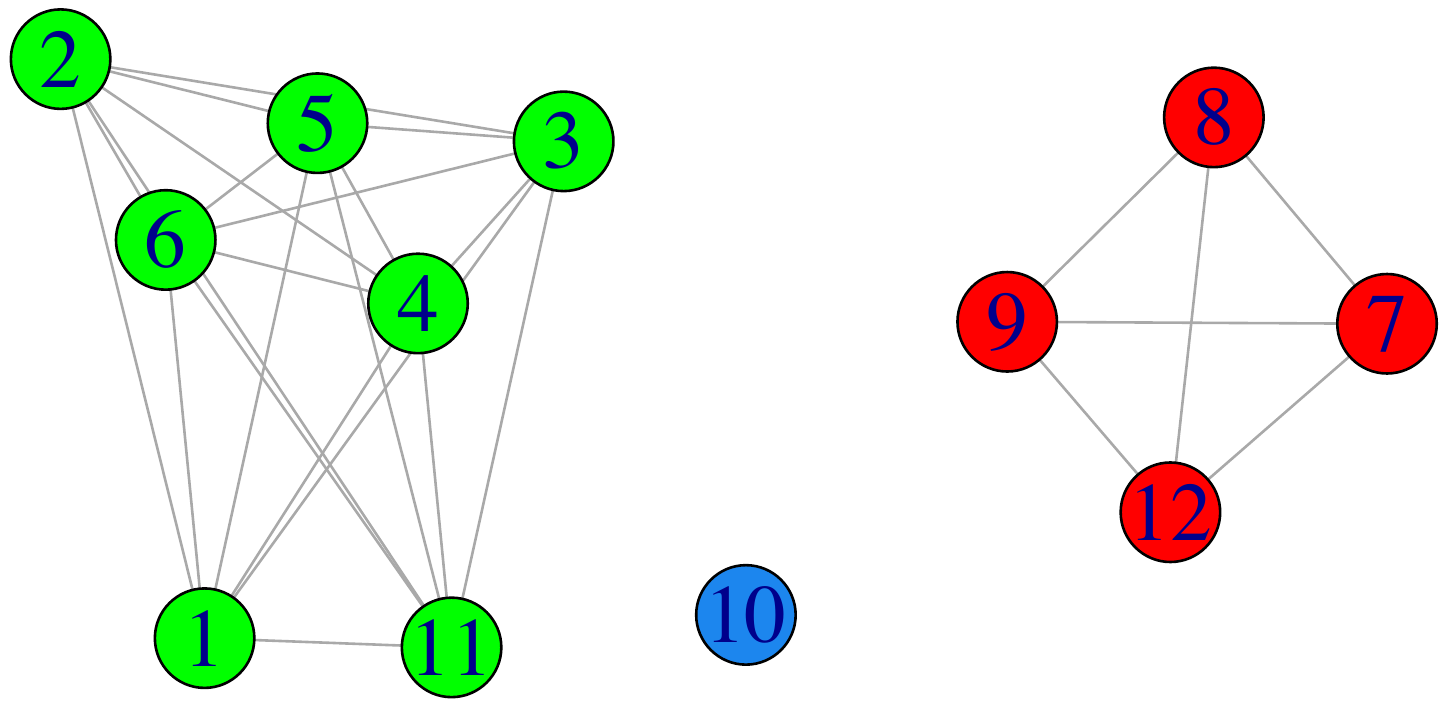}
\caption{12-node influential clique in early phase of election}\label{fig:donor_subgraph_1}
\end{subfigure}\\
\centering
\begin{subfigure}{0.33\textwidth}
\centering
\includegraphics[width=1\textwidth, height = 0.09\textheight]{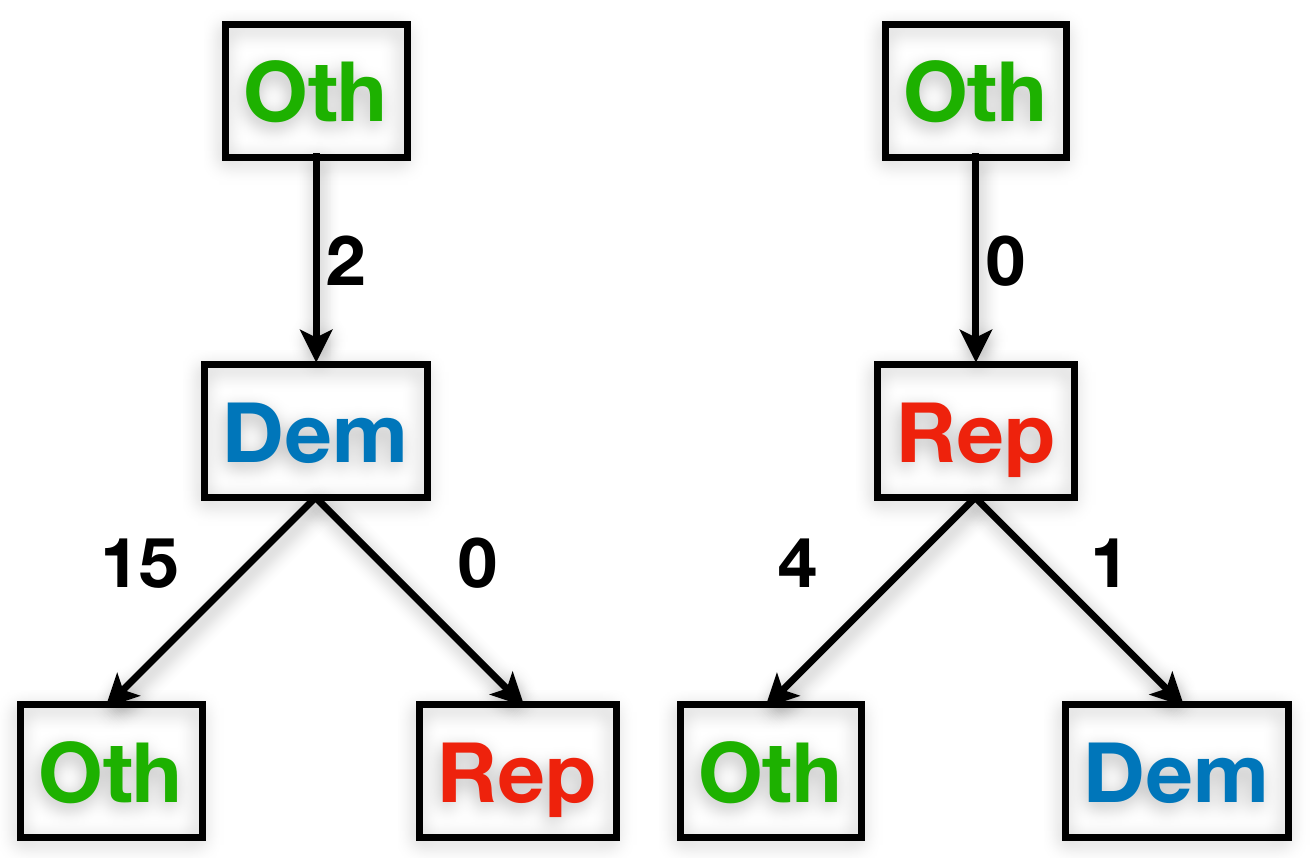}
\caption{Democrats lost more donors than Republicans among the highest donors in the top 100 anomalies}\label{fig:election_trends_top}
\end{subfigure}\hfill
\begin{subfigure}{0.33\textwidth}
\centering
\includegraphics[width=0.8\textwidth]{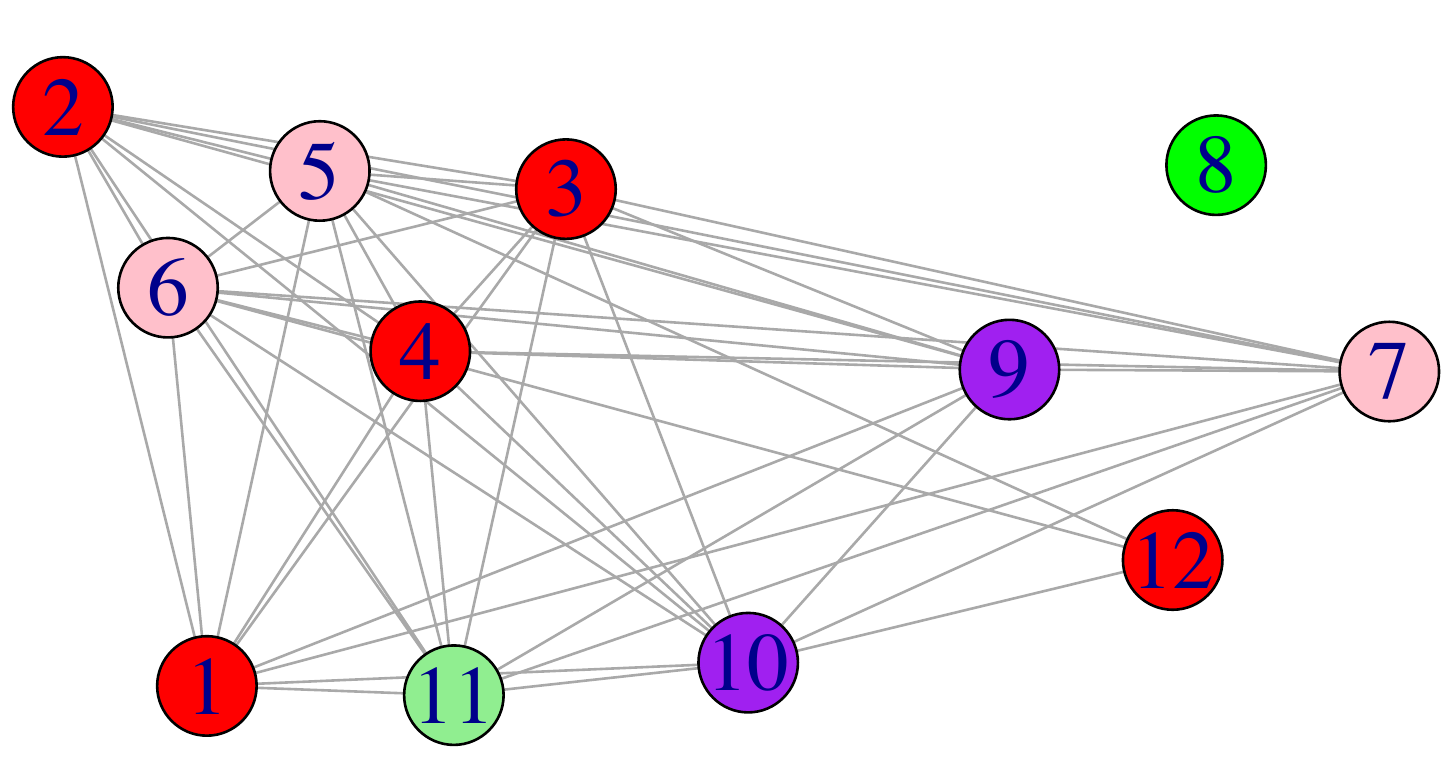}
\caption{The same 12-node influential clique in the final phase of election}\label{fig:donor_subgraph_2}
\end{subfigure}

\caption{Result from analyzing top 100 anomalies in the election data}
\end{figure}

In our election donation analysis, we study two different graphs. In the first graph which has edge weights based on actual donation amounts, the top 100 anomalies are highly skewed towards donors who donated very high amounts to the campaign either in the early or final phase of the election timeline. This indicates a dominating effect of the $|A_1 - A_2|$ term in $\Delta E$ computation (see Algorithm \ref{alg:caddelag}). Some of the important anomalies that got highlighted in this study are Thomas Murphy (Patrick Murphy's father), David Boies, and Henry Howard. Because of large donations (\$2 million) from  Thomas Murphy to \emph{Floridians For A Strong Middle Class}, which exclusively supports his son Patrick Murphy (Democratic senate candidate in 2016 election) and \emph{the Senate Majority PAC}, Patrick's campaign received strong criticism.  Eventually, Patrick was defeated by incumbent Republican Senator Marco Rubio in the general election. 
David Boies, known as `Democratic superlawyer', represented controversial characters like Peter Strzok and Harvey Weinstein (\url{http://bit.ly/2EWLlzT}).
Figure~\ref{fig:election_trends_all} shows the aggregate shifts in donation that happened across party lines from the early phase of the elections to the final phase for the top 100 anomalies. Figure~\ref{fig:election_trends_top} shows the same results for only those donors who donated more than \$10,000 in both phases of the election (and have been identified as anomalous by \name). It can be seen in both figures that the largest volume of shifts happened in donors who donated to the Democratic party during the early phase, but directed their donations to  the independent ``Others'' during the last six months leading to the elections. Although  not a strong enough signal for predicting the election outcome, this discovery made due to \name, makes it possible to identify a general shift in support among the donors, which can act as a potential indicator of the likely election outcome, which most exit polls and data analyses failed to uncover\footnote{Article dated 10.11.2016 in NYT: How Data Failed Us in Calling an Election} during the 2016 presidential elections.

In our second study, to suppress the effect of donation amounts on the final list of anomalies, we considered the edge weights as logarithms of the donation amounts. We also divided the donors into three categories based on donation amounts (see section \ref{data}). In this study our top 100 anomalies and their top contributors consisted of 512 unique individuals, with 12 of them appearing in at least 10\% of the results. Figure \ref{fig:donor_subgraph_1} shows the subgraph structure for those 12 donors in the early phase of the election and Figure \ref{fig:donor_subgraph_1} shows the same subgraph in the final phase of the election. The donors are color coded to reflect their support, with red,  blue, and green indicating Republican, Democrat, and Others respectively. In the first graph there are three distinct subgraphs that are not directly connected to each other. However, in the second graph we see a number of changes in their party support. With the exception of 8 who changes her support to Others, almost everyone ends up supporting overlapping parties (mostly Republicans, as can be seen in the color change of the nodes), thereby significantly changing the graph structure. On further investigation, it is revealed that all of these nodes are linked to small or large businesses either as owners or in executive positions. As the election date approached more people in power positions donated to the Republican party, thereby indicating a change in strategy or preference.

\section{Conclusion and Future Work}
\label{conclusion}

Graph structural anomaly detection algorithms are useful in many applications where understanding relational changes is more useful than independent tuple-based analysis. Ad-hoc sparsification of dense graphs for computational purposes often leads to loss of relevant information. In this paper we propose \name which does commute time based anomaly detection that has been shown to have promise in finding graph structural anomalies \cite{sricharan2014localizing}. The main contribution of this paper is an Apache Spark based distributed framework for computing commute time embedding of very large dense graphs that do not fit in memory. We do this by developing a memory and I/O optimized Spark implementation of block matrix multiplication that does not suffer from the usual scaling issues of Spark based matrix multiplication. We provide thorough theoretical and experimental analysis of \name with respect to accuracy and scalability. We demonstrate the usefulness of \name in identifying anomalies in a climate graph sequence, that have been historically missed due to ad-hoc graph sparsification and on an election donation data set. For future work, we want to extend this framework for other graph structure analysis tasks.

\bibliographystyle{plain}
\bibliography{bibfile.bib}

\end{document}